\pdfoutput=1

\documentclass[journal]{IEEEtran}
\usepackage{graphicx}
\usepackage{subcaption}
\usepackage{cite}
\usepackage{overpic}
\usepackage{setspace}
\usepackage{type1cm}
\usepackage{caption}
\usepackage{comment}
\usepackage{cite}

\usepackage{amsmath}
\usepackage{amsfonts}
\usepackage{mathrsfs}
\usepackage[ruled,linesnumbered]{algorithm2e}  

\usepackage{algorithmic}
\usepackage{multirow} 
\usepackage{multicol} 
\usepackage{arydshln}
\usepackage{colortbl}

\usepackage{array}

\usepackage{fixltx2e}

\usepackage{tikz}
\usetikzlibrary{matrix,positioning,shapes,shadows,arrows,calc,decorations.pathreplacing}

\tikzstyle{blocky} = [draw, line width=1.2pt,fill=white, rectangle, 
minimum height=2.5em, minimum width=4.5em, rounded corners]    

\tikzstyle{blocky2} = [draw, line width=1.2pt,fill=red!10, rectangle, 
minimum height=2.5em, minimum width=18.5em, rounded corners]

\definecolor{darkred1}{RGB}{228,26,28}
\definecolor{darkblue1}{RGB}{55,126,184}
\definecolor{darkgreen1}{RGB}{77,175,74}

\begin{document}
%
\title{RetinaMatch: Efficient Template Matching of Retina Images for Teleophthalmology}
%
%
%

        
\author{Chen Gong$^{\star}$, N. Benjamin Erichson$^{+}$, John P. Kelly$^{\star \dagger}$, Laura Trutoiu$^{\dagger}$, Brian T. Schowengerdt$^{\dagger}$, Steven L. Brunton$^{\star}$, Eric J. Seibel$^{\star}$,        
\thanks{$^{\star}$ Mechanical Engineering, University of Washington, Seattle}%
\thanks{$^{+}$ Department of Applied Mathematics, University of Washington, Seattle}%
\thanks{$^{\star \dagger}$ Department of Ophthalmology, University of Washington, Seattle}%
\thanks{$^{\dagger}$ Magic Leap Inc., FL 33313, U.S.A.}
\thanks{\textbf{This work has been submitted to the IEEE for possible publication. Copyright may be transferred without notice, after which this version may no longer be accessible.}}}

\maketitle

\begin{abstract}
Retinal template matching and registration is an important challenge in teleophthalmology with low-cost imaging devices. However, the images from such devices generally have a small field of view (FOV) and image quality degradations, making matching difficult.    
In this work, we develop an efficient and accurate retinal matching technique that combines dimension reduction and mutual information (MI), called RetinaMatch. The dimension reduction initializes the MI optimization as a coarse localization process, which narrows the optimization domain and avoids local optima.
The effectiveness of RetinaMatch is demonstrated on the open fundus image database STARE with simulated reduced FOV and anticipated degradations, and on retinal images acquired by adapter-based optics attached to a smartphone. RetinaMatch achieves a success rate over 94\% on human retinal images with the matched target registration errors below 2 pixels on average, excluding the observer variability. It outperforms the standard template matching solutions. In the application of measuring vessel diameter repeatedly, single pixel errors are expected. In addition, our method can be used in the process of image mosaicking with area-based registration, providing a robust approach when the feature based methods fail.
To the best of our knowledge, this is the first template matching algorithm for retina images with small template images from unconstrained retinal areas.   
In the context of the emerging mixed reality market, we envision automated retinal image matching and registration methods as transformative for advanced teleophthalmology and long-term retinal monitoring.	
\end{abstract}


\section{Introduction}
\label{sec:intro}

Telemedicine applications are emerging at a rapid pace relying on innovations in hardware and software, and changing attitudes of clinicians, providers and consumers. As an important part in telemedicine, teleophthalmology is now arguably the standard of care in linking patients in remote areas to ophthalmologists. 
Recently, low-cost teleophthalmology has been facilitated by smartphone-based fundus imaging.
In addition, the emerging virtual and mixed reality sector may enable new teleophthalmology scenarios for long-term eye imaging and monitoring. However, in the case of portable fundus photography, non-mydriatic image quality is more vulnerable to distortions, such as uneven illumination, noise, blur and low contrast \cite{wang2016human}.
In this paper, we address the challenging problem of retinal image matching and registration to enable future teleophthalmology applications. 

\subsection{Motivation}
The eye provides a unique opportunity to image internal biological tissue in vivo and many diseases can be diagnosed and monitored through ocular imaging. For example, diabetic retinopathy is a common retinal complication associated with diabetes, causing microaneurysms, exudates and hemorrhages on the retina \cite{shi2015}. Changes of retinal arteries and veins, as well as their ratios, can be indicators of hypertension \cite{Kawasaki2009}. The timely detection of these pathological changes via regular retinal screening and analysis is particularly important for early diagnosis and prevention. 

High quality fundus images of the retina are traditionally acquired in a laboratory setting with expensive and cumbersome equipments. Acquiring high quality fundus images poses a challenge for patients in rural and other underserved areas who must overcome significant hurdles to receive regular checkups in the clinic. Visiting an ophthalmologist often is not convenient for people in the city as well.
In contrast, emerging portable and low-cost fundus cameras allow fast, accessible imaging of the retina, albeit with a decrease in image quality. Using portable fundus cameras outside the clinic connects rural patients with their doctors \cite{panwar2016fundus,Fink2016}. 
By daily retinal monitoring and trend analysis of the data, ocular disease may no longer be considered the silent disease, as early onset is likely to be detectable and even predicted \cite{roesch2017automated}. 


A typical example of such fundus cameras is clip-on lens adapters attached to smart-phone systems \cite{panwar2016fundus}, while these consumer-grade optical devices have two main disadvantages: small FOV and lower image quality than lab-based fundus cameras. The FOVs of current clip-on lenses range from 5$^{\circ}$  to 20$^{\circ}$ in undilated eyes \cite{ludwigfuture, panwar2016fundus}. 
In this case, many small images captured in undilated eye at different locations are necessary to obtain adequate retinal imaging. The same retinal locations need to be re-imaged and matched in order to monitor changes longitudinally over time.
Accordingly, all of the captured small FOV images can be registered and compared to a stored wide FOV retinal image. This reference image is a baseline which can be stitched by a series of small FOV images together, or can leverage wider FOV images captured from a conventional ophthalmoscope. Taking the small FOV images as the templates to be matched, it is a template matching process, as shown in Fig. 1(a).    
The template only covers a small area on the retina, thus is unlikely to be affected much by the nonlinear deformation due to the non-planar eyeball surface. The relationship between the template and the full image can be modeled by linear transformations. We set the problem as the template matching under 2D affine transformations.     

As described above, an accurate template matching method to deal with small FOV and low quality template images is supposed to be come up with for teleophthalmology. Since the method will be implemented on portable devices, the efficiency of computation is also a driving requirement. In the next section, we discuss related prior work on retinal template matching. 


%
\begin{figure*}[!tp]
	\centering
	
	\begin{subfigure}[t]{0.24\textwidth}
		\centering
		\DeclareGraphicsExtensions{.eps}
		\includegraphics[width=1\textwidth]{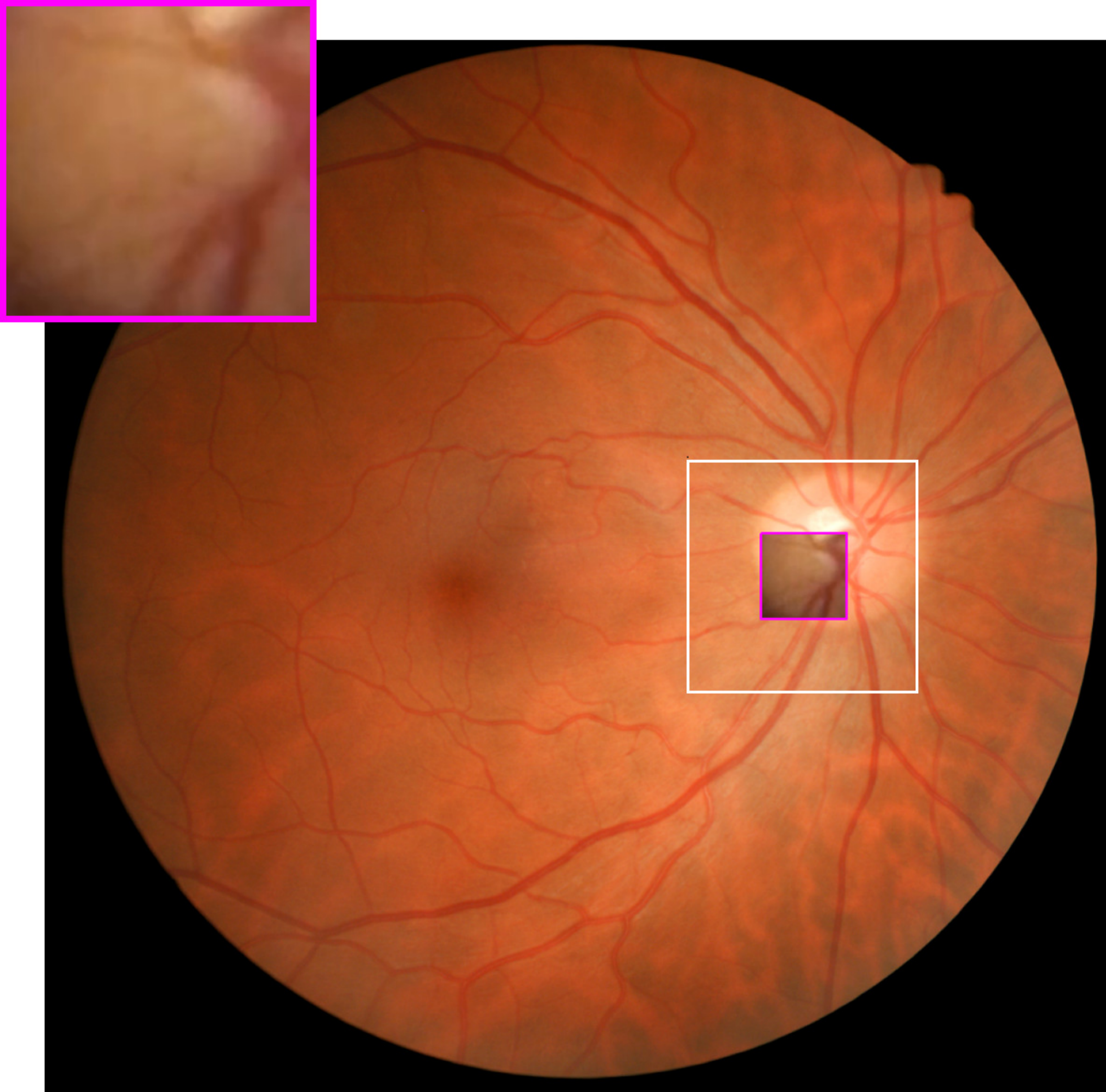}
		\caption{Template matching sample}
	\end{subfigure}
	~
	\begin{subfigure}[t]{0.23\textwidth}
		\centering
		\DeclareGraphicsExtensions{.eps}
		\includegraphics[width=1\textwidth]{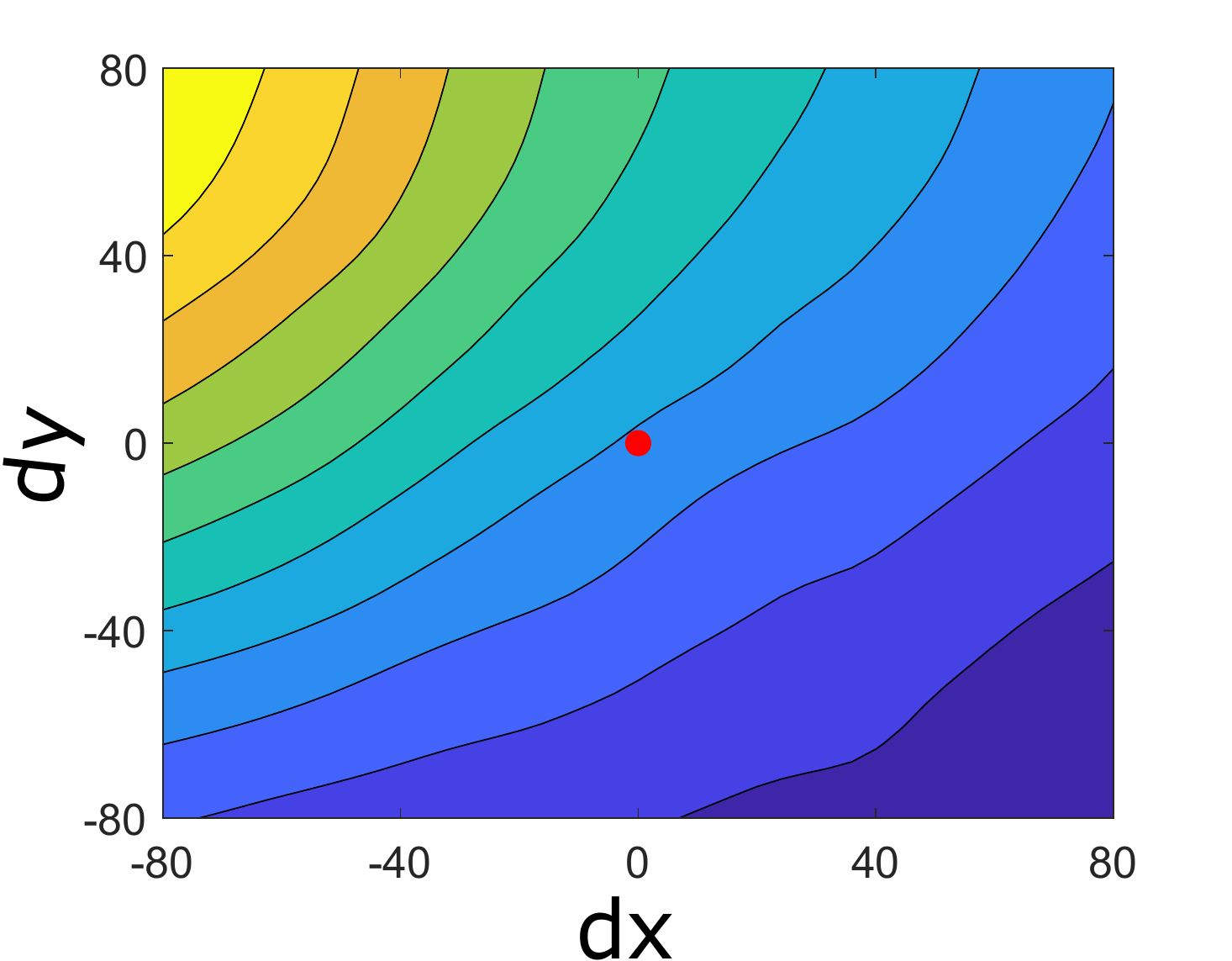}
		\caption{SSD}
	\end{subfigure}
	~
	\begin{subfigure}[t]{0.23\textwidth}
		\centering
		\DeclareGraphicsExtensions{.eps}
		\includegraphics[width=1\textwidth]{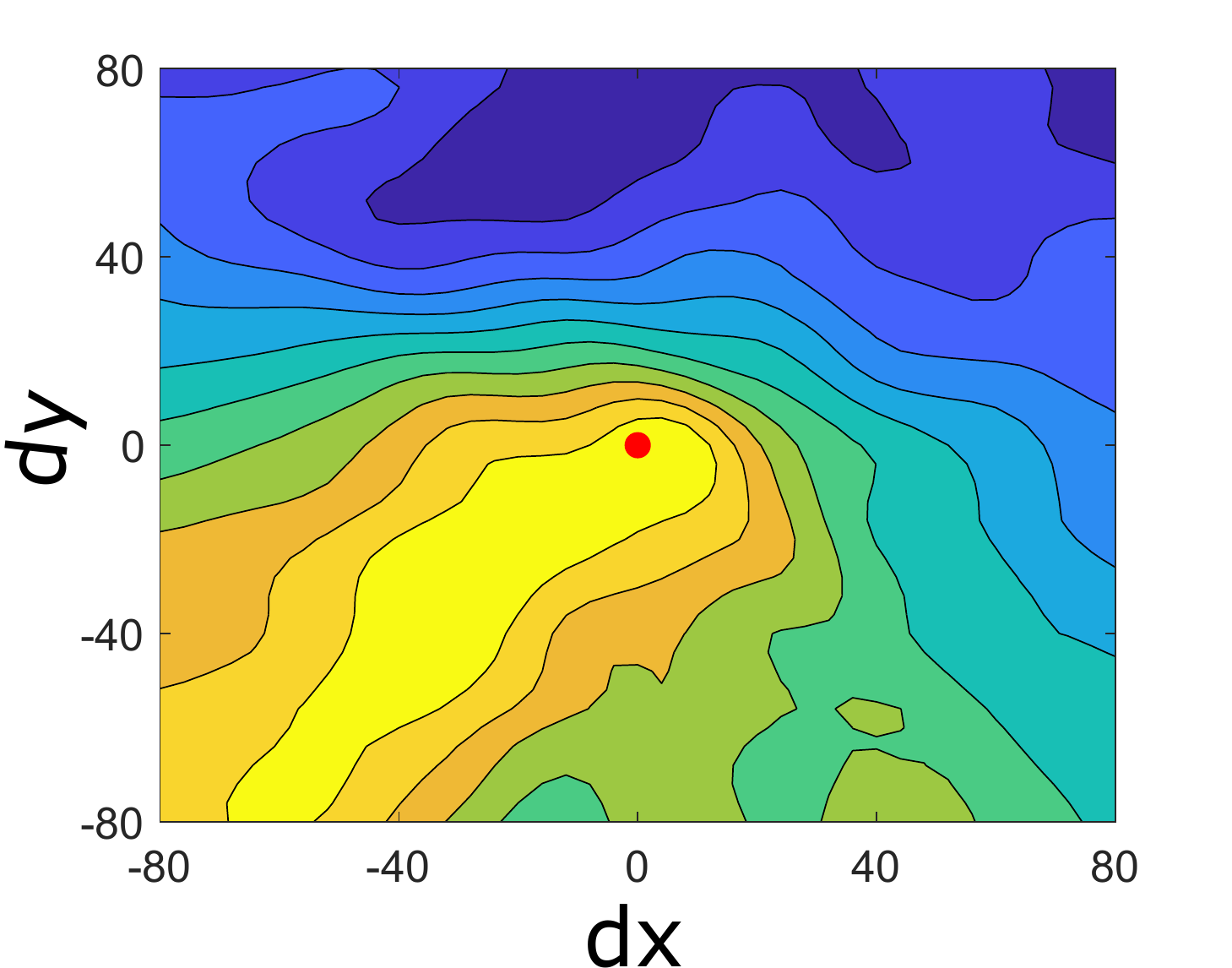}
		\caption{NCC}
	\end{subfigure}
	~
	\begin{subfigure}[t]{0.23\textwidth}
		\centering
		\DeclareGraphicsExtensions{.eps}
		\includegraphics[width=1\textwidth]{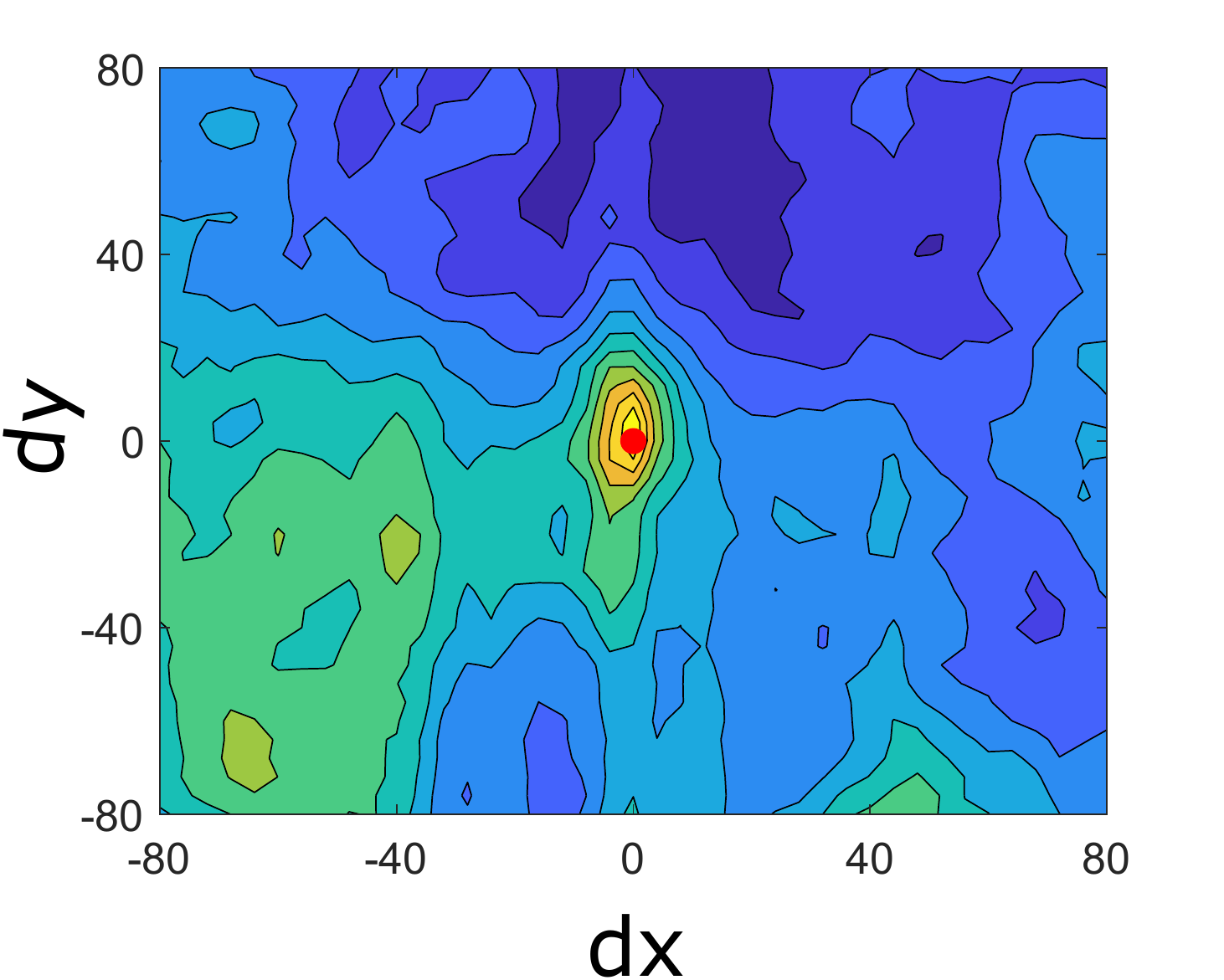}
		\caption{NMI}
	\end{subfigure}		
	
	\caption{Alignment functions with respect to translations between the template and the white boxed area. The full FOV image in (a) is taken with the fundus camera in the clinic. The top left image in the magenta square is the template captured by a typical adapter-based fundus camera \textit{D-eye}, having only translations along two axes. 
	(b)(c)(d) show the alignment function between the template and regions within the white boxed area. The true alignment position is $(0,0)$ -- see red dots. Only NMI shows an obvious maximum at the alignment position. Note that the optimal value of SSD is minimum and NCC and NMI are maximums. }
	\label{fig:1}
\end{figure*}
\subsection{Related Work}


Much of the foundational work on template matching of retinal images is based on more general image registration methods, which have been comprehensively studied in recent years. However, general retina registration methods focus on matching image pairs that both have a large FOV with local deformations or different image modalities. The existing retinal template matching algorithms are limited to detecting specific objects from the image, where the template always contains a certain feature, such as the optic disc, exudate and artifacts \cite{yu2012fast,zhang2014exudate,mora2013template}.  

Retinal image registration itself is challenging: the nonvascular surface of retina is homogeneous in healthy retinas, while exhibiting
a variety of pathologies in unhealthy retinas \cite{stewart2003dual}. Retina images captured by adapter-based optics provide less information and have low image quality, which further increases the difficulty of template matching. It is instructive to introduce current retina image registration methods which can be used for template matching and their feasibility in addressing our stated problem.
Retina image registration approaches can be classified into area-based and feature-based methods. Feature based methods optimize the correspondence between extracted salient objects in retina images \cite{sofka2006retinal,stewart2003dual,xu2007optic}. Typically, bifurcations, fovea, and the optic disc are common features used for retinal image registration. A small FOV template has little probability of containing specific landmarks on the retina, thus the fovea and optic disc are not applicable. Vascular bifurcations are more common, while similarly, the small amount of bifurcations in the template cannot form the basis of a robust registration. Besides, the extraction of the vascular network in poor quality images is difficult. It can cause ambiguous vascular directions when label the bifurcations. 
General feature point based approaches are also implemented in retina registration, such as SIFT-based \cite{wang2010automatic,tsai2010edge} and SURF-based methods \cite{wang2015robust,hernandez2015retinal}. These approaches can register the images in complex scenarios and are computationally efficient. They assume the feature point pairs can be reliably detected and matched to estimate the transformation. Although feasible in most cases, the process can fail on low-quality retina images without enough distinct features.

Area-based approaches match the intensity differences of an image pair under a similarity measure, such as SSD (sum of squared differences) \cite{friston1995spatial}, CC (Cross-Correlation) \cite{cideciyan1995registration} and MI (mutual information) \cite{zhu2007mutual}, then optimize the similarity measure by searching in the transformation space. Avoiding pixel-level feature detection, such approaches are more robust to poor quality images than feature-based approaches. However,  retina images with sparse features and similar backgrounds are likely to lead the optimization into local extrema. 
Fig. \ref{fig:1} shows an example of the area-based method with three similarity measures. The small template image is captured by the adapter-based optics \textit{D-eye} which is registered onto a full fundus image. Both of the images are acquired by the same modality. SSD and normalized CC (NCC) do not have an obvious peak at the alignment position (0,0), giving no clear information on the alignment quality. Normalized MI (NMI) shows a maximum at the alignment position, while it still has local extrema which can interfere the global optimization. Besides, when the size difference between the template and full image is too large, the registration with MI can be computationally very expensive.

\subsection{Contributions}
In this paper, we present RetinaMatch, a new template matching method that overcomes the challenges posed by registering small FOV and low-quality retinal images onto a full image.
This approach is an improvement over the area-based method with MI metric, since it is more reliable to achieve accurate and robust template matching near the alignment position, as shown in Fig. \ref{fig:1}. The unique aspect of our approach is that we combine dimension reduction methods with the MI-based registration to reduce the inference of local extrema and improve the matching efficiency. An overview of the general idea is shown in Fig. \ref{fig:overview}.
Specifically, the principal component analysis (PCA) and block PCA are used to localize the template image coarsely, then the resulting displacement parameters are used to initialize the MI metric optimization. The initial parameters provided by the coarse localization are in the convergence domain of MI metric. In this way, the transformation search space for optimization is narrowed significantly. The PCA computation is accelerated with randomized methods, which improves the coarse localization efficiency. Another contribution is that this paper proposes an efficient image mosaicking algorithm based on the image dimension reduction. It accelerates the matching of overlapped images among unordered data, especially in the image mosaicking with area-based registration methods.

The proposed method is validated on the STARE retinal dataset \cite{STARE} with synthetic deformations, and the \textit{in vivo} data captured by a low-cost ($<$US\$400) adapter-based optical device \textit{D-eye}. The performance of different dimension reduction techniques are also compared on the STARE dataset. RetinaMatch can find the correct mapping even when the image is of poor quality with non-distinct features; whereas other compared methods fail due to unstable feature detection and local extrema interference.

\section{Preliminaries}
\subsection{PCA for Location Estimation}


Dimension reduction methods allow the construction of low-dimensional summaries, while eliminating redundancies and noise in the data. To estimate the template location in the 2d space, the full image dimension is redundancy, thus we apply dimension reduction methods for the template coarse localization. In this section we describe the dimension reduction methods we used in this paper.      

Generally, we can distinguish between linear and nonlinear dimension reduction techniques. The most prominent linear technique is principal component analysis (PCA), which dates back to the work of~\cite{pearson1901liii} and~\cite{hotelling1933analysis}. PCA is selected as the dimension reduction method in RetinaMatch since it is simple and versatile.
Specifically, PCA forms a set of new variables as a weighted linear combination of the input variables.
Consider a matrix $\mathbf{X}=[\mathbf{x}_1,\mathbf{x}_2,...,\mathbf{x}_d]$ of dimension $n\times d$, where $n$ denotes the number of observations and $d$ is the number of variables. Further, we assume that the matrix $\mathbf{X}$ is column-wise mean centered. The idea of PCA is to form a set of uncorrelated new variables (so called principal components) as a linear combination of the input variables:
\begin{equation}
\mathbf{z}_i = \mathbf{X}\mathbf{w}_i,
\end{equation}
where $\mathbf{z}_i$ is the $i$th principal component (PC) and $\mathbf{w}_i$ is the weight vector. The first PC explains most of the variation in the data, the subsequent PCs then account for the remaining variation in descending order. Thereby, PCA imposes the constraint that the weight vectors are orthogonal. This problem can be expressed compactly as the following minimization problem:
\begin{equation}\label{eq:pca_min}
\begin{aligned}
& \underset{}{\text{minimize}}
& & \|\mathbf{X-ZW}\|_F^2 \\
& \text{subject to}
& & {\bf W^\top W = I},
\end{aligned}
\end{equation}
where $\|.\|_F$ is the Frobenius norm. The weight matrix $\mathbf{W}$ that maps the input data to a subspace turns out to be the right singular vectors of the input matrix $\mathbf{X}$. Often a low-rank approximation is desirable, i.e., we compute only the $k$ dominant PCs using a truncated weight matrix $\mathbf{W}_k=[\mathbf{w}_1,\mathbf{w}_2,...,\mathbf{w}_k]$. 

PCA is generally computed by the singular value decomposition (SVD). Many algorithms have been developed to streamline the computation of the SVD and PCA for high-dimensional data that exhibits low-dimensional patterns \cite{kutz2016dynamic}. In particular, tremendous strides have been made to accelerate the SVD and related computations using randomized methods for linear algebra~\cite{halko2011finding,erichson2018randomized,erichson2018sparse}. 
Since we have demonstrated high performance with less than 20 principal components, the randomized SVD is used to compute the principal components, improving the efficiency in this retinal mapping application for mobile platforms.
The randomized algorithm proceeds by forming a sketch $\mathbf{Y}$ of the input matrix 
\begin{equation}
\mathbf{Y} = \mathbf{X} \mathbf{\Omega},
\end{equation}
where $\mathbf{\Omega}$ is a $d\times l$ random test matrix, say with independent and identically distributed random standard normal entries. Thus, the $l$ columns of $\mathbf{Y}$ are formed as a randomly weighted linear combination of the columns of the input matrix, providing a basis for the column space of $\mathbf{X}$. Note, that $l$ is chosen to be slightly larger than the desired number of principal components. Next, we form an orthonormal basis $\mathbf{Q}$ using the QR decomposition $\mathbf{Y} = \mathbf{QR}$. Now, we use this basis matrix to project the input data matrix to low-dimensional space
\begin{equation}
\mathbf{B} = \mathbf{Q}^\top \mathbf{X}.
\end{equation}
This smaller matrix $\mathbf{B}$ of dimension $l \times d$ can then be used to efficiently compute the low-rank SVD and subsequently the dominant principal components. Given the SVD of $\mathbf{B} = \mathbf{U}\mathbf{\Sigma}\mathbf{V}^\top$, we obtain the approximate principal components as
\begin{equation}
\mathbf{Z} = \mathbf{Q} \mathbf{U}\mathbf{\Sigma} = \mathbf{X} \mathbf{V}.
\end{equation}
Here, $\mathbf{U}$ and $\mathbf{V}$ are the left and right singular vectors and the diagonal elements of $\mathbf{\Sigma}$ are the corresponding singular values.

The approximation accuracy can be controlled via additional oversampling and power iterations, for details see~\cite{Erichson2016arxivA}. 

\subsection{Mutual Information}

In this section, we describe the maximization of MI for
multimodal image registration. 
We define images $\bf S$ and $\bf \widehat{S}$ as the template and target images, respectively. A transform $u$ is defined to map pixel locations $x \in \bf S$ to pixel locations in $\bf \widehat{S}$. 

The main idea of the registration is to find a deformation $\widehat{u}$ at each pixel location $x$ that maximizes the MI between the deformed template image $ {\bf S}(u(x)) $ and the target image $\bf {\widehat{S}}(x)$. Accordingly,
\begin{equation}
u_{opt}=\mathop{\arg\min}_{u} MI({\bf S}(u(x)),{\bf \widehat{S}}(x)),
\end{equation}
where
\begin{equation}
MI({\bf S}(u(x)),{\bf \widehat{S}}(x))=\sum_{i_1\in S} \sum_{i_2\in \widehat{S}} p(i_1,i_2)log(\dfrac{p(i_1,i_2)}{p(i_1)p(i_2)}).
\label{eq:MI}
\end{equation}
Here, $i_{1}$ and $i_{2}$ are the image intensity values in ${\bf S}(u(x))$
and $\bf \widehat{S}(x)$, respectively, and $p(i_1)$ and $p(i_2)$ are their
marginal probability distributions while $p(i_1,i_2)$ is their joint probability distribution. 

\begin{figure*}[!tp]
	\centering
	\includegraphics[width=0.85\textwidth]{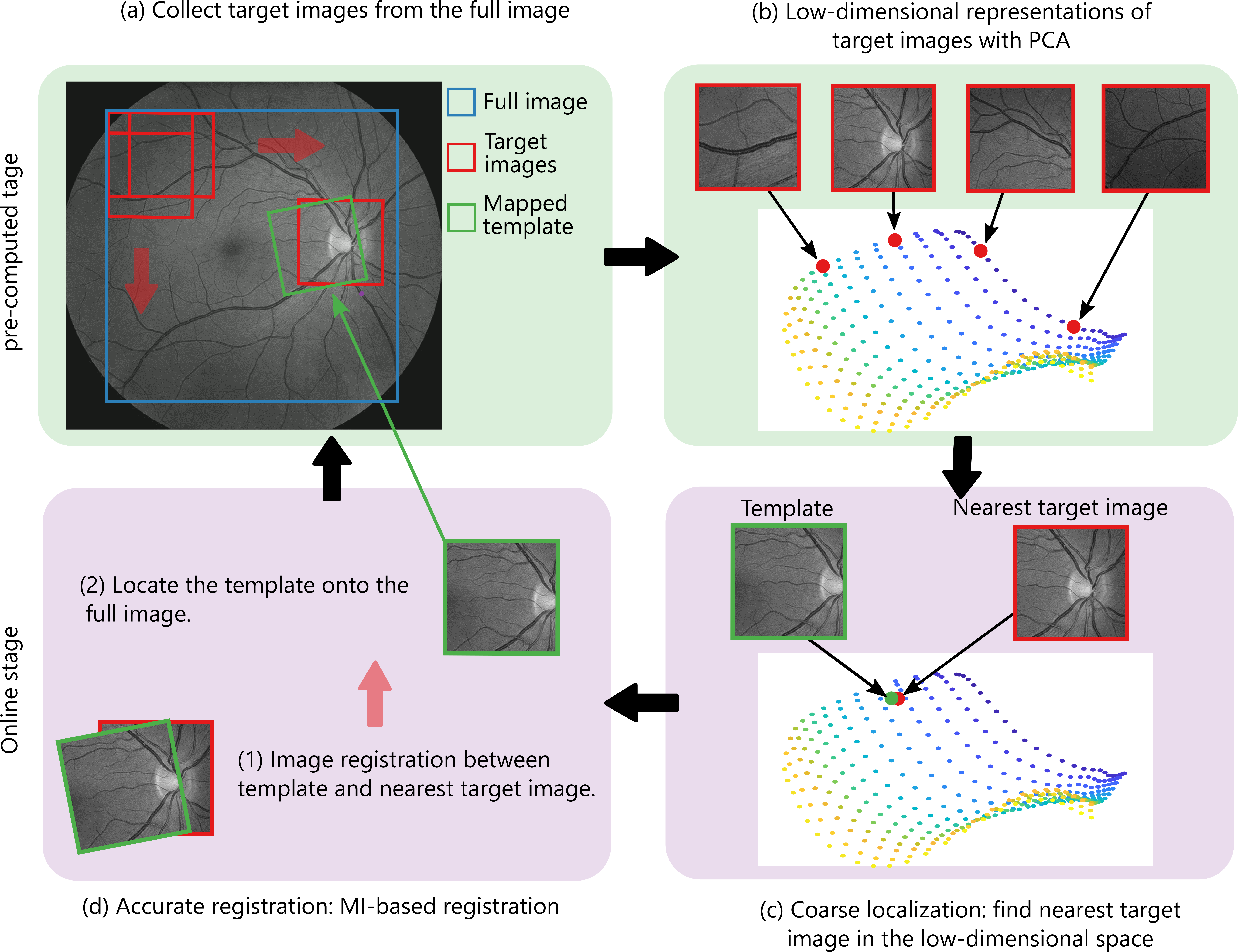}
	\caption{Schematic of the proposed retinal template matching method shown in four panels from (a) to (d) . In panel (a) the wide-FOV full image is sampled with many overlapping target images. (b) Each target image is mapped into the low-dimensional space according to its positional relationship. (c) An example template is also mapped into this space and its nearest target image is found. (d) The nearest target image is registered with MI. The panels (a) and (b) in green can be pre-processed offline when the full image is obtained, while panels (c) and (d) are considered as the online stage. The schematic describes the method without using the improvement of block PCA. Please see Fig. \ref{fig:block PCA} in Sect. III for more details of block PCA.}
	\label{fig:overview}
\end{figure*}

\section{Proposed Approach}

In this section, we describe the RetinaMatch combining dimensionality reduction and mutual information based image registration. From Fig. \ref{fig:1} we can see MI performs better than other similarity metrics even on the same modality images, thus we focus on the MI criterion. Given a large FOV full image and a small FOV template image, our method can be used to localize the template on the full image accurately and efficiently. The full image can be a wide-field fundus image or a mosaicked one from D-eye images. The underlying concept is to use PCA and block PCA first for coarse localization, which can be a warm start to follow accurate registration. In accurate registration, the MI metric is optimized to find the optimal transformation.   
Since the optimization domain has been narrowed to a small range near the optimal position with coarse localization, the accurate registration can achieve high accuracy and efficiency. Fig. \ref{fig:overview} provides an overview of the general approach to RetinaMatch.


\subsection{Coarse Localization with Dimension Reduction}


We define the full image and the template as $\bf F$ and $\bf S$ respectively. The full image $\bf F$ is split into target images $ {\bf I}_{1}, {\bf I}_{2},...,{\bf I}_{N}$:
\begin{equation}
 {\bf I}_{i} = \phi(b_{i},\bf F). 
\end{equation}
The function $\phi$ crops ${\bf I}_{i}$ from $\bf F$ at $b_{i}$ and $b_{i} = [x_{i}, y_{i}, h, w]$, where $ (x_{i}, y_{i}) $ denotes the center position and $ (h, w) $ denotes the width and height of the source image. There is a certain displacement $f$ of neighboring target images in the $x$ and $y$ axes. As shown in Fig. \ref{fig:overview}(a), each target image has a large overlap with its neighbors. The overlap forms the redundancy of the data which can indicate the location distribution between each image and its neighbors. Applying dimension reduction techniques on such data we can obtain the low-dimensional distribution map of all target images.
 
Target images are resized to vectors and form the matrix $\bf X$ $\in \mathbb{R}^{n \times d}$. We obtain the low-dimensional distribution representation of the target image distribution by implementing PCA on $\bf T$:
\begin{equation}
{\bf Z}= {\bf X}{\bf W},
\end{equation}
where ${\bf Z}=[z_{1}, z_{2}, z_{3},...,z_{N}]^{T} \in \mathbb{R}^{n \times l}$, ${\bf W} \in \mathbb{R}^{d \times l}$ and $l\ll d$. The image space $\Omega_{1}$ is mapped to a low-dimensional space $\Omega_{2}$ with the mapping $\bf W$.   $\bf W$ and $\bf Z$ are saved in the dictionary $ \mathcal{D}$.

Given a template $\bf S$, the coarse position can be estimated by recognizing its nearest target image. The nearest target image in $\Omega_{1}$ should also be the nearest representation of $\bf S$ in $\Omega_{2}$. Accordingly, we obtain the low-dimension feature $z_{s}$ of the template in $\Omega_{2}$:
	\begin{equation}
	 z_{s}= \tilde{s} \bf W, 
	\end{equation}
where $\tilde s \in \mathbb{R}^{d}$ is the reshaped vector of template $\bf S$. Let $ \Delta(z_{s},  {z}) $ be the Euclidean distance between $z_{s}$ and a feature ${z}$ in $\bf {Z}$. $ {z}^{*} $ is the nearest target feature of the source image $ \bf S $ in $\Omega_{2}$:
\begin{equation}
 {z}^{*} = \mathop{\arg\min}_{{z}} \Delta(z_{s},  {z}).
\end{equation}
The corresponding target image location is used as the coarse location of $ \bf S$. Ideally, the difference between the coarse location and the ground truth in $x$ and $y$ axes should be less than $f/2$ pixels. 

In the first experiment in Sect. IV, PCA outperforms other non-linear dimension reduction methods, while the error is larger than $f/2$. The main reason is that the image degradation creates spurious features that contribute to the final classification. To reduce the influence of local spurious features, we implement block PCA to further improve the accuracy of the coarse localization. By computing the PCA of different local patches in the template, the effect of local features, which leading to the tempate can not be located correctly, is reduced.       
\begin{figure}[!tp]
	\vspace{-0.4cm}
	\centering
	\includegraphics[width=0.40\textwidth]{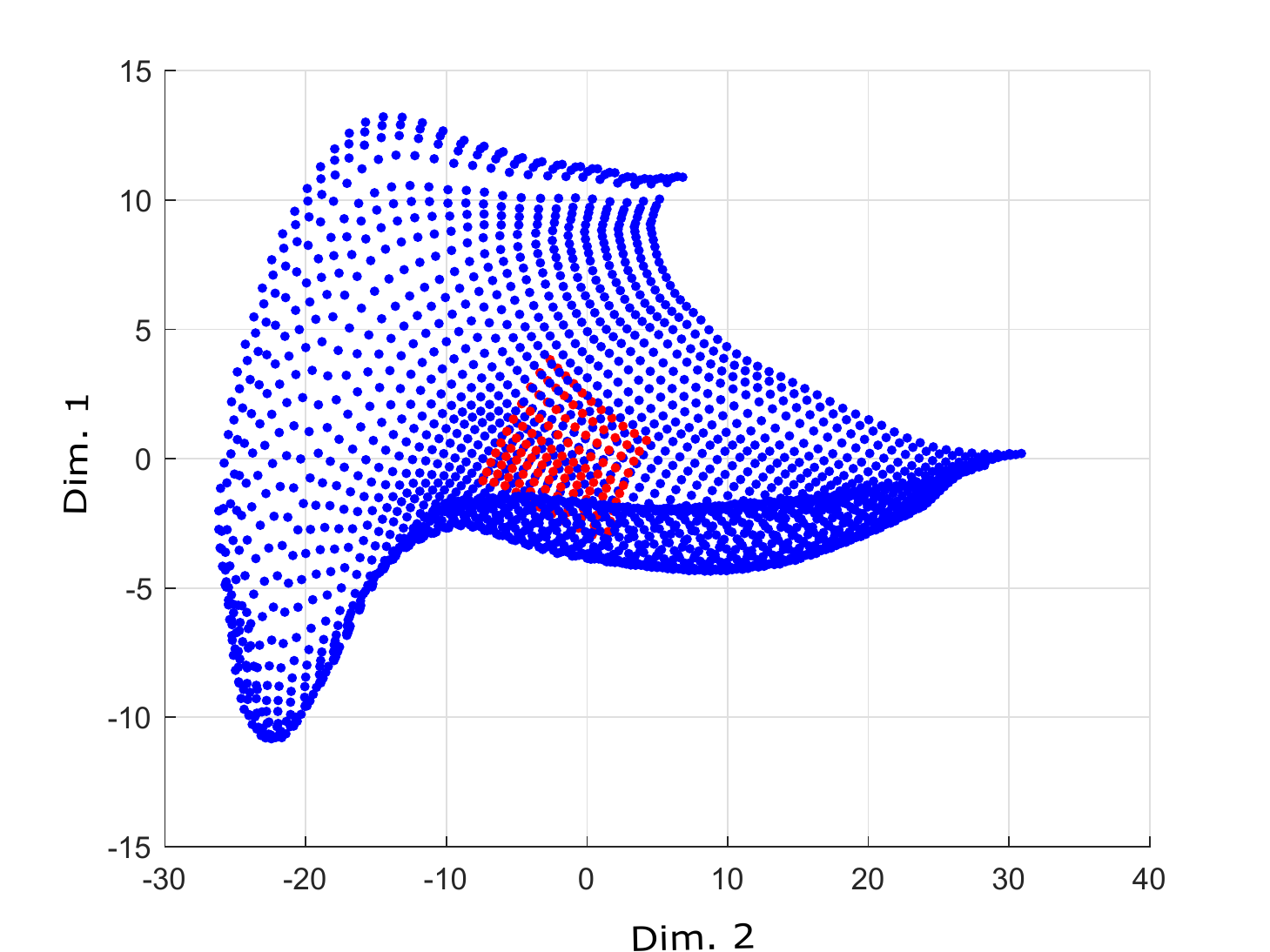}
	\caption{Low-dimension representation of block PCA: Showing only the first two dimensions, each dot denotes one image patch. Blue dots are target  patches and red dots are template patches.}
	\label{fig:block PCA}
	\vspace{-0.4cm}
\end{figure}

Obtaining the nearest target image, we crop a larger image at the same position from the full image as the new target image $\bf I$. In this way, the template can have more overlap with the new target image when there is a large offset between two images. We segment $\bf I$ and the template $\bf S$ into small patches with the function $\tilde{\phi}$, where the patch size is smaller than the source image with the axial displacement of neighboring patches $f'$. Similarly, all image patches from $\bf I$ are mapped into the low-dimension space $\Omega_{3}$ with $ \bf W'$. Let $\bf Z'$ denote the low-dimensional representation of the target image distribution. Each template patch is then mapped to the space with $ \bf W'$. As shown in Fig. \ref{fig:block PCA}, all blue dots denote the target image patches and red dots denote the template patches, which can indicate the positional relationship between two images. 
The nearest target patch for each template patch is determined with the Euclidean distance as described before. The coordinates of each target patch represent the locaion of the mapped patch.
We use the same weight for each region of the template for localization, thus the average of all template patches location can be taken as the template's location.  Let $b_m$ be the mean of the coordinates of selected nearest target patches, which then represents the center of the template on $\bf I$. Accordingly, the template location on the full image can be estimated and the region is cropped as the image $\bf \widehat{S}$. The accurate registration is then applied to the template $\bf S$ and image $\bf \widehat{S}$. In this way, the coarse localization provides an estimate of a good initial point for the accurate registration.

In the implementation of the proposed coarse localization, the full image is assumed to exist so the dictionary $\mathcal{D}$ can be built in advance. This is the pre-computed part as shown in Fig. \ref{fig:overview} (a-b). The process after the template being acquired is called the online stage, involving the block PCA for coarse localization followed by the accurate registration. The online stage of the coarse localization is shown in Algorithm 1.

\begin{algorithm}
	\SetAlgoLined
	Map template $\bf S$ into space $\Omega_{2}$: $z_{s}=\tilde{s} \bf {W}.$
	
	Determine closest target image $\bf I$ with corresponding ${z}^{*}$ : ${z}^{*} = \mathop{\arg\min}_{{z}} \Delta(z_{s},  {z}).$  $z^* \in \bf Z$.
	
	Segment $\bf S$ into $[S_p^{1}, S_p^{2},...,S_p^{n}]$: $S_p^{i} = \tilde{\phi} (b_{i},\bf S) $; Segment $\bf I$ into $[I_p^{1}, I_p^{2},...,I_p^{n}]$: $I_p^{i} = \tilde{\phi} (b_{i},\bf I) $.
	
	Map target patches $I_p^{i}$ into space $\Omega_{3}$: ${\bf Z'}= {\bf I_p}{\bf W'}$, where $\bf I_p$ is formed with vectorized $I_p^{i}$.
	
	For each template patch $S_p^{i}$:
	
	\quad (i)Map $S_p^{i}$ into space $\Omega_{3}$: $ \tilde{z_s}^{i}=S_p^{i} \bf W'$.
	
	\quad (ii)Determine its closest target patch $I_p^{Idx(i)}$ with \quad index $Idx(i)$.  
	
	$b_m=\dfrac{1}{n} \sum\limits_{i=1}^{n} b_{Idx(i)}$, where $b_{Idx(i)}$ is the coordinate of selected target patch $I_p^{Idx(i)}$.
	
	
	\Return localization region $\widehat{S} = \phi(b_m,\bf F))$.
	
	\caption{Coarse localization: online stage}
\end{algorithm}  

 \subsection{Accurate Registration}
 
 In this section, images $\bf S$ and $\bf \widehat{S}$ are accurately registered with maximization of mutual information. The location of $\bf \widehat{S}$ on the full image $\bf F$ becomes the estimated displacement of the template $\bf S$. As the small FOV of template images, the relationship between the template and the full image can be modeled by linear transformations. In our work, the transform $u$ for alignment is given as an affine transformation:
 \begin{equation}
 u= \left[
 \begin{matrix}
 a_{11} & a_{12} & t_{x} \\
 a_{21} & a_{22} & t_{y} \\
 1 & 1 & 0
 \end{matrix}
 \right].
 \end{equation}
 
 From the MI equation \ref{eq:MI}, we can see the MI function has a discrete formulation which is not differentiable. Several solutions therefore are proposed to smooth the MI function to compute the MI derivatives and keep its accuracy. We use the method described in \cite{thevenaz1997spline}, where
 the joint probability distribution between the images $\bf S$ and $\bf \widehat{S}$ is estimated with a Parzen window. 
 
 The optimizer used for the MI maximization is based on Newton's method. The MI function is a quasi-concave function (Fig. \ref{fig:1}(d)), and the parabolic hypothesis of the Newton’s method is only valid near the convergence. When the initial tranformation on the convex part of the cost function, it will cause the optimization to diverge. In the example of Fig. \ref{fig:1}(d), the normalized MI measure has local extrema interference. The proposed coarse localization provides a good initialization of the displacement for subsequent optimization of the MI cost function. In the figure, the estimated alignment position is $(11,9)$. The estimation is close to the optimal value and falls in the convex domain of the MI metric, which provides more efficient optimization and avoid local extrema. 
 
 After registration between images $\bf S$ and $\bf \widehat{S}$, the template $\bf S$ can be matched on the full image $\bf F$ based on the position $b_m$ of the selected region $\bf \widehat{S}$.

\subsection{Image Stitching}

As pointed out in Sect. I, the full retina image can be stitched into a panorama by using many small templates. 
Users must capture a series of images in naturally unconstrained eye positions to explore different regions of the retina.
It is problematic to determine adjacent images before the registration when we apply area-based registration approaches, since they do not have effective descriptors for matching. 
 

Related to the dimension reduction in the proposed template matching method, here we present Algorithm 2 to learn the positional relationship of images to be stitched. In this way, the adjacent images can be recognized and registered efficiently. For a series of small images $\bf X_{i}$, we form the matrix $\bf X$, as with the matrix $\bf T$. PCA is applied to $\bf X$ and returns the low-dimensional features for each image in  $\Omega_{2}$. The distance between features in $\Omega_{2}$ indicates the distance between images. The nearest neighbor $\bf X_{j}$ of image $\bf X_{i}$ is the one with largest overlap; the image pair is then registered with MI-based approach. To improve the algorithm robustness, the $3$-nearest neighbors for each image are first selected to compute MI with, and we keep the one with the largest metric value. 


\begin{algorithm}
	\SetAlgoLined
	Map images into space $\Omega_{2}$:${\bf Z}= {\bf X}{\bf W}.$
	
	For each image $\bf X_{i}$:
	
	\quad (i).Find the nearest $3$ neighbors $\bf X_{j}$ minimizing the feature distance $ \Delta(\bf Z_{i},  Z_{j}) .$ 
	
	\quad (ii).Compute the Mutual Information between each $\bf X_{j}$ and $\bf X_{i}$ and take the adjacent image with highest MI.
	
	Panorama $\bf R$ Mosaicking: Align all the adjacent images with mutual information based registration method.
	
	Panorama blending.

	\Return panorama $\bf R$.
	
	\caption{Image stitching}
\end{algorithm}

\section{Experiments}
\label{sec:EXPERIMENTS}

We present the performance of our template matching method on three experiments using retina images. The proposed algorithm is implemented in Matlab. For comparison, we use the global MI algorithm described in Mattes et al. \cite{mattes2003pet} and ASIFT (modified SIFT for affine deformation) \cite{morel2009asift}. In the first experiment, each template is extracted from the full fundus image in the STARE dataset and matched back to it. The intermediate results of the coarse localization are also evaluated. In the second experiment, the template captured by the adapter-based optics is matched to the full fundus image captured by the clinical fundus camera. In the third experiment, a panorama is mosaicked from small templates first, and subsequently individual templates are matched to the panorama. 


\subsection{Fundus Images from STARE Dataset}

In this experiment, we validate our method on simulated fundus images. We use images from the STARE dataset \cite{STARE}, which consists of 400 raw fundus images of healthy and diseased retinas. Matching image pairs are simulated from this dataset. Each image pair includes a full fundus image selected randomly from the dataset and an affine transformation is applied to map it from a square into a parallelogram. The area within the mapped square is then cropped and warped (with the inverse affine transformation) to obtain the square template. The FOV of the template images is around 12$^{\circ}$ with a size of $200\times 200$ pixels. The template dimension is 10\% of the full image. 

The ground truth is available in this experiment, thus root-mean-square (RMS) errors between corrected displacements and ground truth positions are used as a metric to measure the RetinaMatch accuracy.
To evaluate the coarse localization, we take the center point distance between the template and the chosen target region.

\subsubsection{Validation of the Coarse Localization}

First the coarse localization with and without the block PCA refinement are tested. In the implementation, target images are generated with a displacement of $f=10$ pixels and $f'=5$ for the block PCA. We use the top 20 and 10 PCA features in the first PCA step and the block PCA respectively. The parameters are fixed in remaining experiments. Additionally, we test the coarse localization with two other non-linear dimension reduction methods: kernel PCA \cite{scholkopf1997kernel} and Isomap \cite{tenenbaum2000global}. We compare the non-linear dimension reduction methods to see if the non-planar retina surface and the affine transformation affect the performance of the PCA-based linear method. The experiment is performed over 100 matching image pairs created from the STARE dataset. The pixel-level errors (coarse localization error as described), success rates, and average runtimes of these methods are shown in Table I. The criterion of successful matches is a pixel-level error of less than 40 pixels.  
It is verified that the PCA based coarse localization is more efficient, accurate and interpretable. Block PCA further improves the accuracy while the time spent is higher than PCA-only method. To further improve the online efficiency, the target patches mapping can be precomputed for each target images. The average time spent in this case will decrease to 0.0975s.

\begin{figure}[!tp]
	\centering
	
	\begin{subfigure}[t]{0.14\textwidth}
		\centering
		\DeclareGraphicsExtensions{.eps}
		\includegraphics[width=1.07\textwidth]{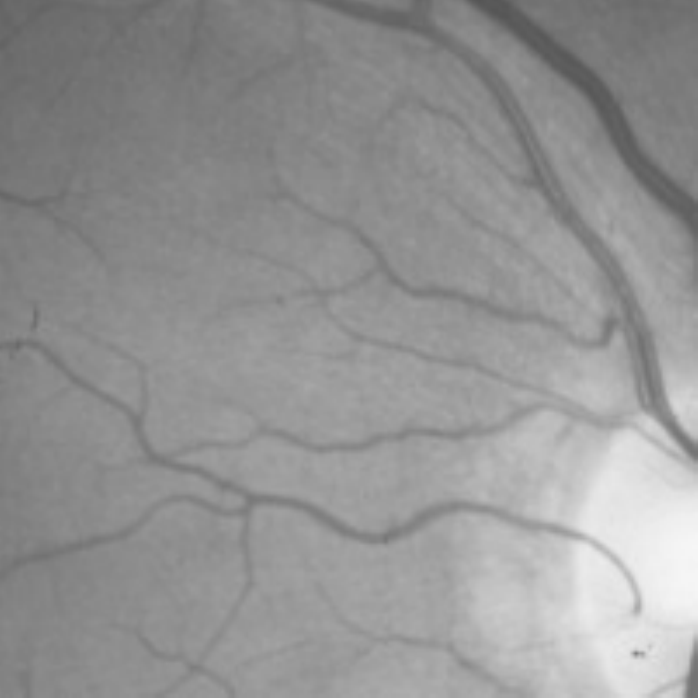}
		\caption{Normal}
	\end{subfigure}
	~
	\begin{subfigure}[t]{0.14\textwidth}
		\centering
		\DeclareGraphicsExtensions{.eps}
		\includegraphics[width=1.07\textwidth]{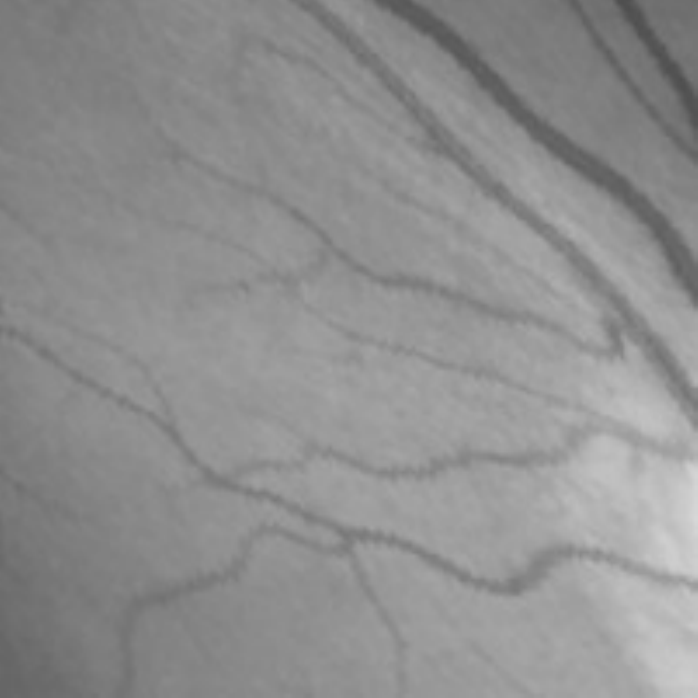}
		\caption{Affine deformation}
	\end{subfigure}
	~
	\begin{subfigure}[t]{0.14\textwidth}
		\centering
		\DeclareGraphicsExtensions{.eps}
		\includegraphics[width=1.07\textwidth]{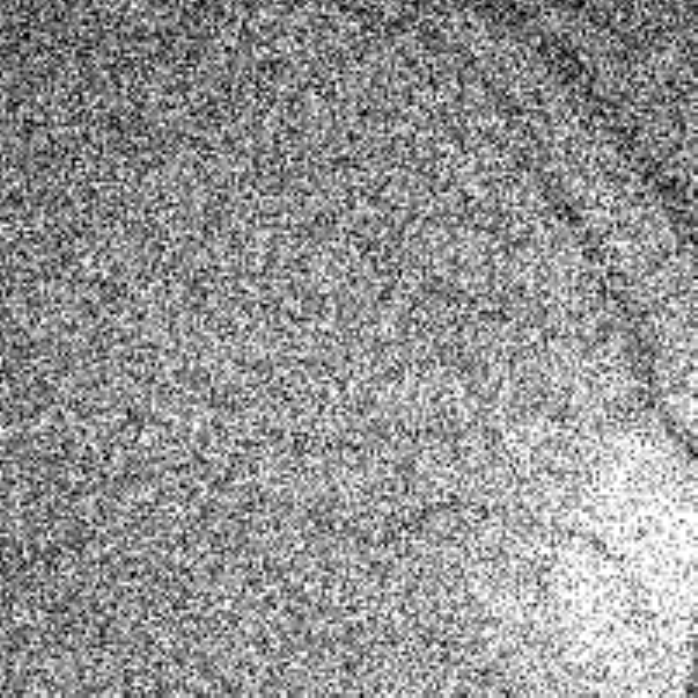}
		\caption{Gaussian noise}
	\end{subfigure}
	
	\begin{subfigure}[t]{0.14\textwidth}
		\centering
		\DeclareGraphicsExtensions{.eps}
		\includegraphics[width=1.07\textwidth]{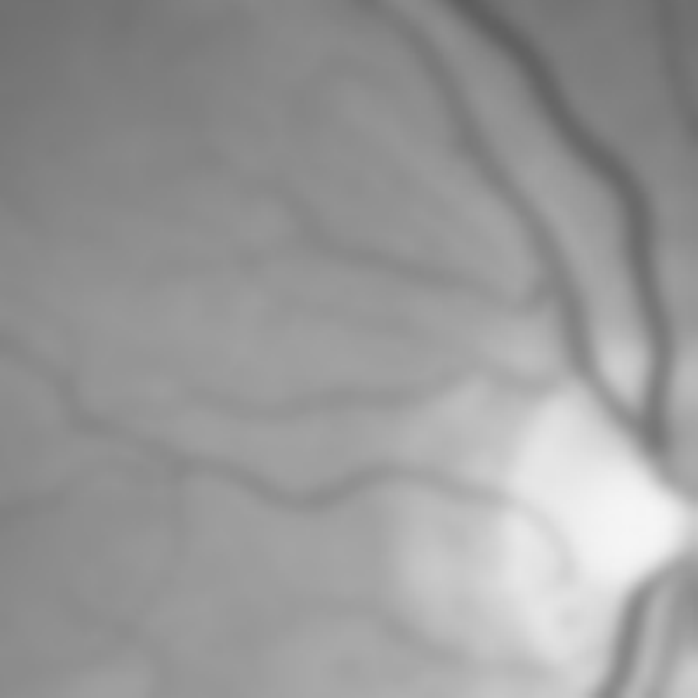}
		\caption{Blur}
	\end{subfigure}	
	~
	\begin{subfigure}[t]{0.14\textwidth}
		\centering
		\DeclareGraphicsExtensions{.eps}
		\includegraphics[width=1.07\textwidth]{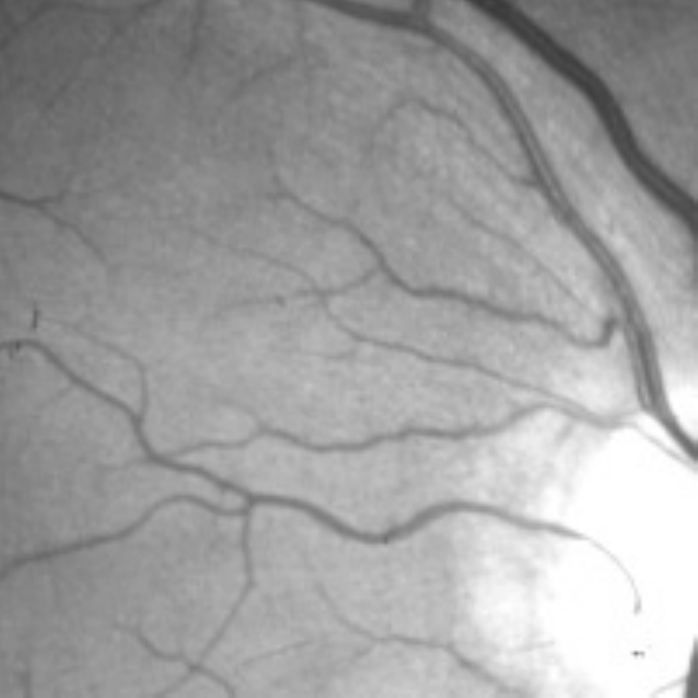}
		\caption{Brightness}
	\end{subfigure}	
	~
	\begin{subfigure}[t]{0.14\textwidth}
		\centering
		\DeclareGraphicsExtensions{.eps}
		\includegraphics[width=1.07\textwidth]{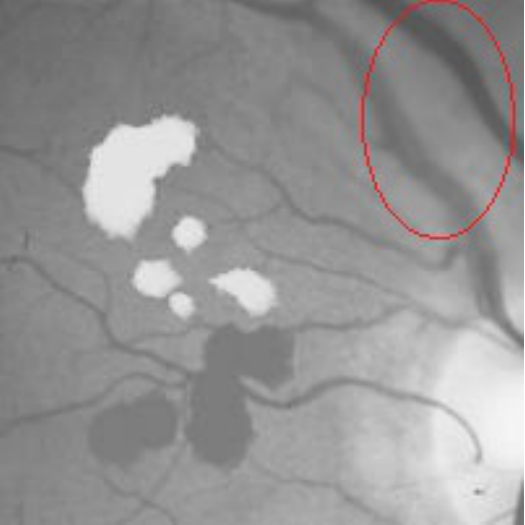}
		\caption{Artificial features}
	\end{subfigure}		
	
	\caption{Examples of highest level degradations in each sequence. Please note that (b) is the template generated with affine deformation thus the image content is not the same. In the artificial features (f), the bright and dark spots simulate the exudate and hemorrhage, respectively. The width of vessels in the circled area is enlarged.}
	\label{fig:degradations}
\end{figure}

Additional experiments were carried out to test the proposed coarse localization under different image degradations. Five degradation types in five levels are considered as follows (images are in double format $ \in [0, 1]$): affine transform with the rotation/shear parameter of \{5$^{\circ}$/0.1,10$^{\circ}$/0.2,15$^{\circ}$/0.2,15$^{\circ}$/0.3,20$^{\circ}$/0.3\}; additive Gaussian noise with standard deviation varied from 10\% to 50\% of the pixel value; image blurring with Gaussian kernels with standard deviation of \{0.5,1,1.5,2,2.5\}; intensity changes of \{4\%,8\%,12\%,16\%,20\%\} of graylevels in the image, which is the nonlinear brightness change; add artificial pathological features of 1-5 levels with increasing amount and size, such as the spot of exudate (bright spots), hemorrhage (dark spots) and vessel width changes (enlarge/shrink vessel regions).    
For each sequence and degradation level, we create 100 matching image pairs as described above. All degradations are applied to the template in each pair. Fig. \ref{fig:degradations} shows template examples of the highest degradation in each sequence. The coarse localization achieves high success rates across the dataset in different degradations, with the exception of the highest level of linear deformation sequence. Checking a large set of data, we find the real affine deformation in the adapter-based optics imaging is less than level three (rotation/shear: 15$^{\circ}$/0.2). 


\begin{table}
	\centering
	\caption{Comparison of coarse localization with different dimension reduction methods.}	
	\setlength{\tabcolsep}{0.008\textwidth}{
	\begin{tabular}{|c|c|c|c|}
		\hline  
		&Mean errors&Success rate&Runtime\\
		\hline  
		Kernel PCA&57&83\%&0.7035s\\
		\hline
		Isomap&27&94\%&2.3634s\\
		\hline
		PCA&14&100\%&0.0065s\\
		\hline
		Block PCA&8&100\%&0.6143s\\
		\hline
	\end{tabular}}
	\label{T1}
\end{table}

\begin{table}
	\centering
	\caption { Success rates of coarse localization per degradation level.}	
	\setlength{\tabcolsep}{0.008\textwidth}{
	\begin{tabular}{|c|c|c|c|c|c|}
		\hline  
		Distortion level&1&2&3&4&5\\
		\hline  
		Affine deformation&100\%&99\%&95\%&81\%&75\%\\
		\hline
		Noise&100\%&100\%&100\%&100\%&100\%\\
		\hline
		Blur&100\%&100\%&100\%&100\%&100\%\\
		\hline
		Brightness change&100\%&100\%&100\%&100\%&97\%\\
		\hline
		Artificial features &100\%&100\%&100\%&100\%&98\%\\
		\hline
	\end{tabular}}
	\label{T2}
\end{table}

\begin{figure*}[!tp]
	\centering
	\begin{subfigure}[t]{0.3\textwidth}
		\centering
		\DeclareGraphicsExtensions{.eps}
		\includegraphics[width=1\textwidth]{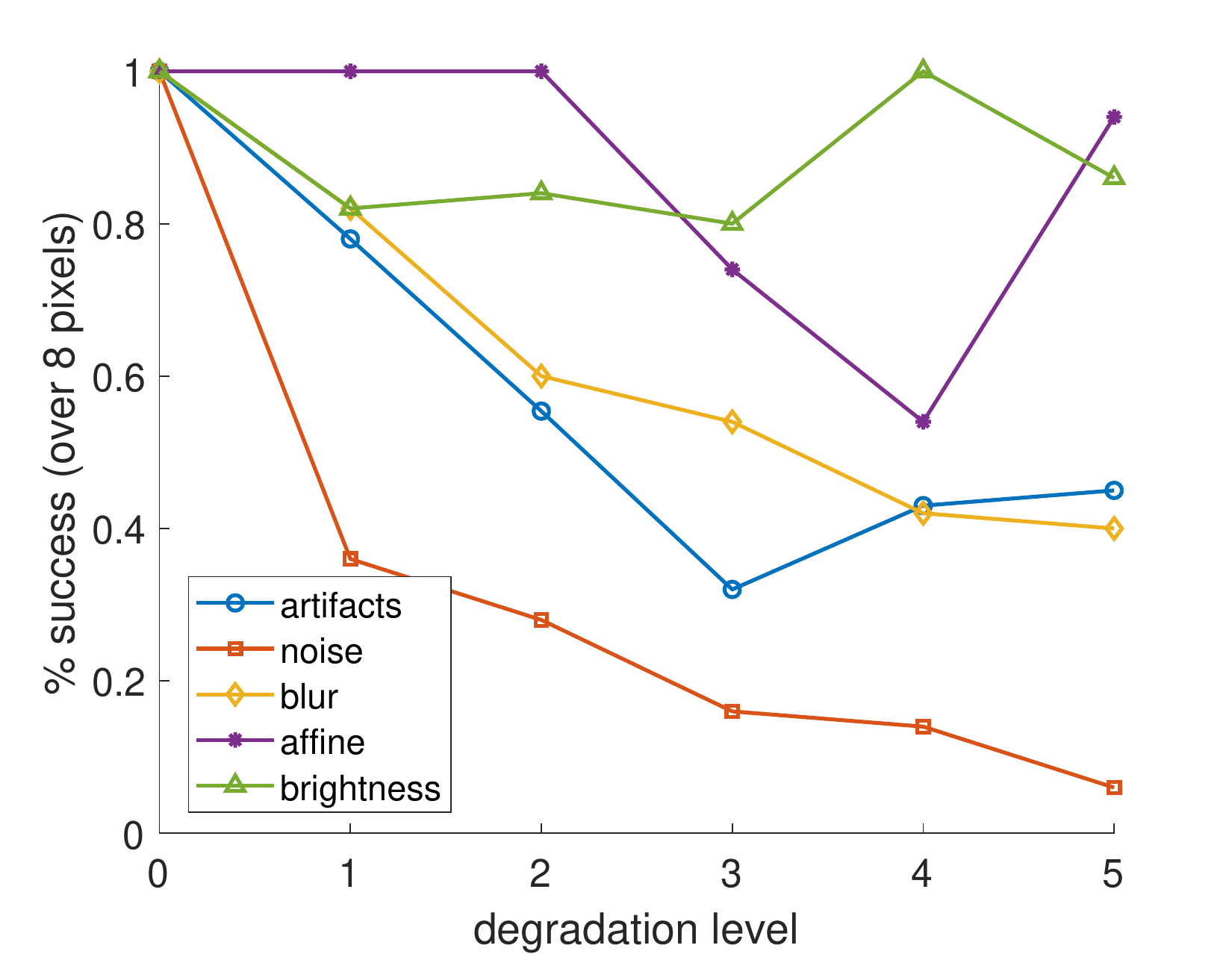}
		\caption{ASIFT}
	\end{subfigure}
	~
	\begin{subfigure}[t]{0.3\textwidth}
		\centering
		\DeclareGraphicsExtensions{.eps}
		\includegraphics[width=1\textwidth]{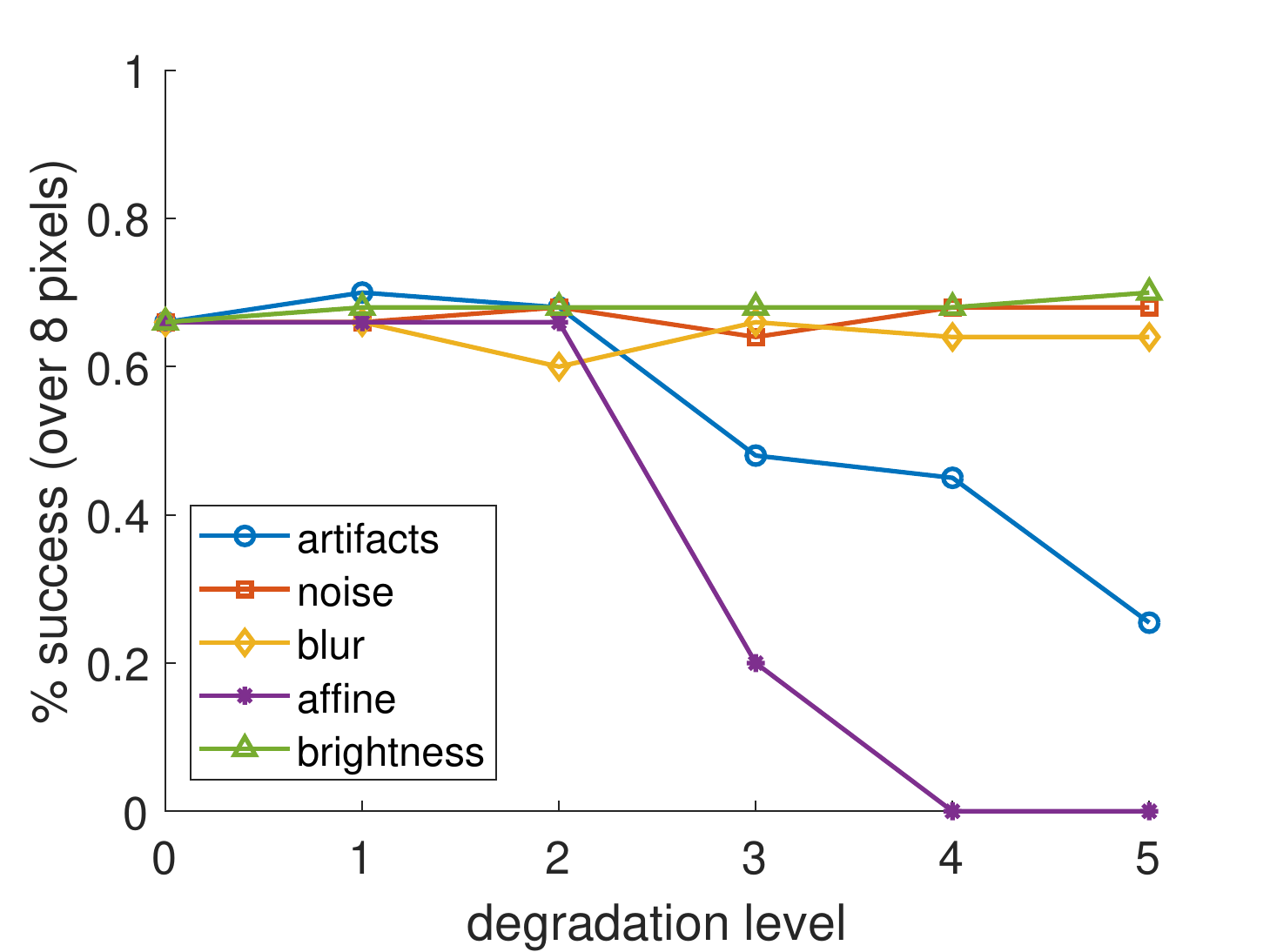}
		\caption{Global MI}
	\end{subfigure}
	~
	\begin{subfigure}[t]{0.3\textwidth}
		\centering
		\DeclareGraphicsExtensions{.eps}
		\includegraphics[width=1\textwidth]{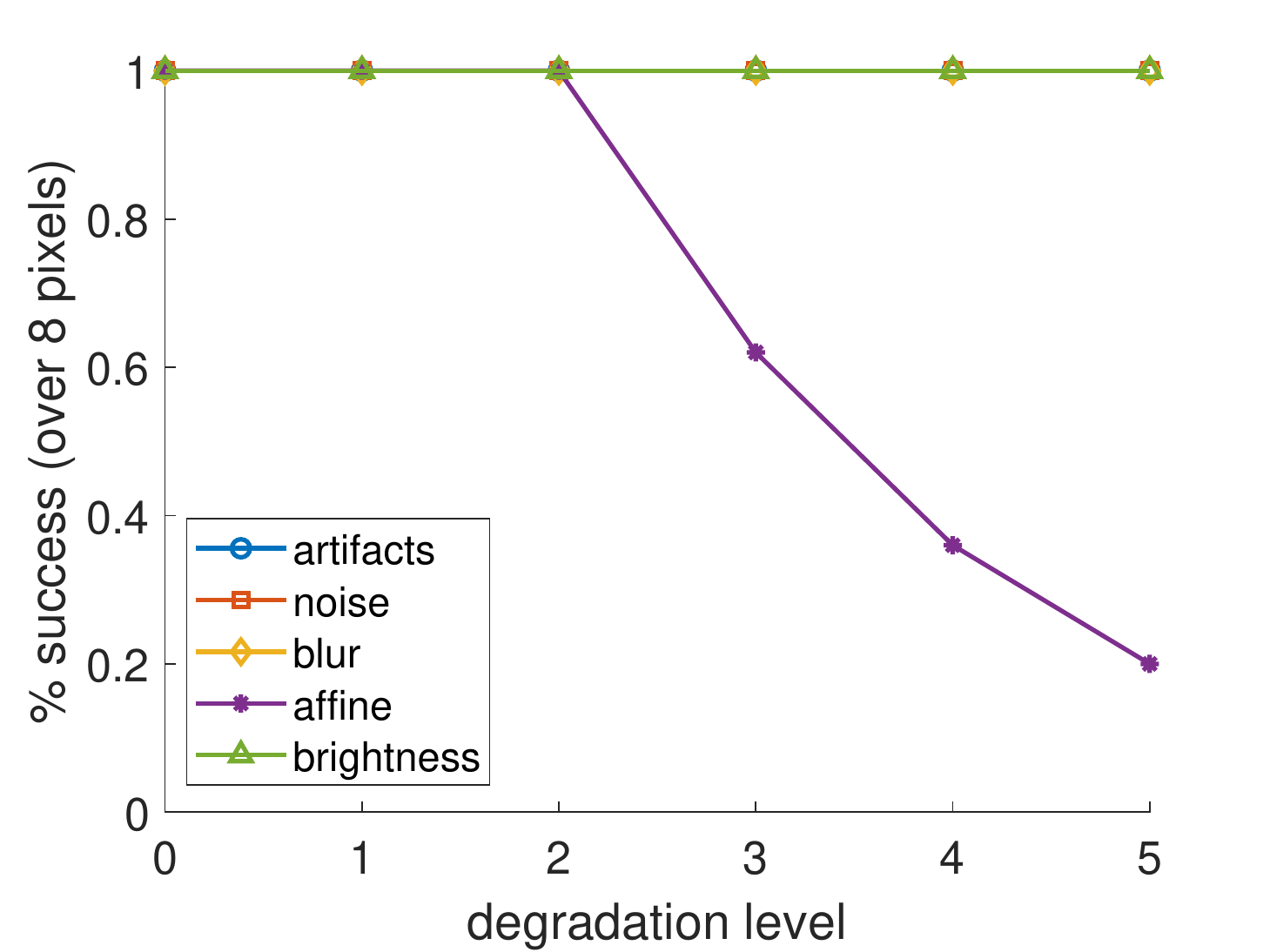}
		\caption{RetinaMatch}
	\end{subfigure}
	\caption{ Performance of template matching methods under different image degradations.  In each, the x-axis stands for the increasing levels of image degradation, ranging from 0 (no degradation) to 5 (highest). The y-axis stands for the percentage of successful matches with RMS error less than 8 pixels. All degradations have 100\% success rate in RetinaMatch except three high-level affine deformations.}
	\label{fig:RM1}
\end{figure*}

\subsubsection{Validation of the Template Matching}     

We examine RetinaMatch's final performance under the same sequences and degradations described earlier, but with two additional template matching approaches: feature-based ASIFT and global MI registration. The RMS errors of mapped pixel coordinates are presented in Fig. \ref{fig:RM1}. 
The accuracy of ASIFT decreases significantly at higher degradation levels of noise, blur, and artifacts, due to feature-point instability.
The global MI registration method cannot always converge to the correct affine transformation parameters using such small templates, thus it has a low success rate even without degradations. The performance further declines in high-level degradations of artifacts and affine. 
RetinaMatch has a success rate of 100\% in most sequences and degradation levels, except the high-level affine deformation. As described above, the real-world affine deformation would be small and not decrease the performance of RetinaMatch. The imrpovement of RetinaMatch efficiency over global MI depends on the size difference between two matched images. The average runtime of RetinaMatch is 50\% of the global MI here. 

\subsection{In vivo D-eye Data and Full Fundus Image}

D-eye is a typical adapter-based optical system which can convert the digital camera on smartphones to a fundus camera (https://www.d-eyecare.com/). Fig. \ref{fig:mosiacPCA} show several examples of D-eye images. 
The relatively small FOV of D-eye can be useful to monitor the retinal health over time with comparison to a wide FOV baseline image taken at the ophthalmology clinic. With our algorithmic approach, the captured D-eye images can be matched onto the full image for automatic comparison. The latest data with retina changes can also replace the original area on the full image, maintaining a record of longitudinal changes. In this way, it offers the opportunity for a quick overall retina analysis outside the clinic, with automatic diagnostic approaches such as described in \cite{walter2002contribution,gulshan2016development}.

This experiment is a case study with a series of D-eye data captured from one person with a healthy retina. We converted the iphone 6 to a fundus camera with D-eye, then collected the data in a dim room to provide a larger pupil and proper image contrast. The eyeball was free to rotate which allowed us to obtain images covering different regions of the retina. The collected \textit{D-eye} images have an average FOV of 4$^{\circ}$ and a resolution of about 50 pixels/degree. The full fundus image is taken with a Kowa Nonmyd alpha-D III retinal camera, as shown in Fig. 1(a). It has a 45$^\circ$ FOV with a resolution of 75 pixels/degree. The D-eye images are around 0.7\% of the full image. Captured with different devices, the brightness and contrast varies greatly between the image pair to be matched. 

We first validate our method by matching 100 \textit{D-eye} template images onto the full image. The ASIFT and global MI methods are also implemented. Additionally, we add pathological artificial features on the 100 D-eye templates to test the algorithm robustness to retina pathological changes.
The accuracy of the template matching is evaluated using target registration error (TRE) \cite{fitzpatrick1998predicting}. For each template, four corresponding landmarks are selected by an trained observer, two trained observers then selected the corresponding landmarks from the full image independently. To obtain TRE for each image pair, we compute the root mean square of the distance between the transformed landmark points and the landmarks observed by trained users.
The TRE results of RetinaMatch (coarse localization and final results), ASIFT and global MI are shown in Table \ref{TRE_exp2}. Table \ref{TRE_exp2} lists the success rate, the mean and standard deviation of TRE of successful matches and inter-observer variability. The success rate is the percentage of successful matching pairs with TRE less than 6 pixels. RetinaMatch can reach an accuracy of less than 4-pixel TRE with the observer variability, while the ASIFT and global MI cannot match the \textit{D-eye} image successfully. 
  

\begin{table*}[tp]
	
	\centering
	\caption{Target registration error (TRE) of template matching methods in experiment 2.}
	\label{TRE_exp2}
	\setlength{\tabcolsep}{0.008\textwidth}{
	\begin{tabular}{|c|c|c|c|c|c|c|c|}
		\hline
	\multirow{2}{*}{}&
\multicolumn{2}{c|}{ASIFT}&\multicolumn{2}{c|}{ MI}&\multicolumn{2}{c|}{RetinaMatch}&{Observer}\cr\cline{2-7}
&Mean$\pm$SD&Success Rate&Mean$\pm$SD&Success Rate&Mean$\pm$SD&Success Rate&Variability\cr
		\hline
		\hline
		Without artificial features& NA&0  & NA& 0& 3.88$\pm$1.72&94\%&2.39$\pm$1.93\cr\hline
		With artificial features& NA & 0 &NA & 0& 3.97$\pm$1.64&94\%&2.39$\pm$1.93\cr\hline
	
		\hline
	\end{tabular}}
\end{table*}

\subsection{In vivo D-eye Data and Mosaicked Full Image}

In this experiment we match the D-eye templates onto the full image mosaicked with D-eye images. Using the stitched panoramic image allows the use of this device at home without going to the clinic for the full fundus image as the baseline. Inhabitants of remote areas without local eye clinics having professional fundus camera facilities can benefit greatly from this technique.

\subsubsection{Full Image Mosaicking }
\begin{figure}[!tp]
	\centering
	\includegraphics[width=0.5\textwidth]{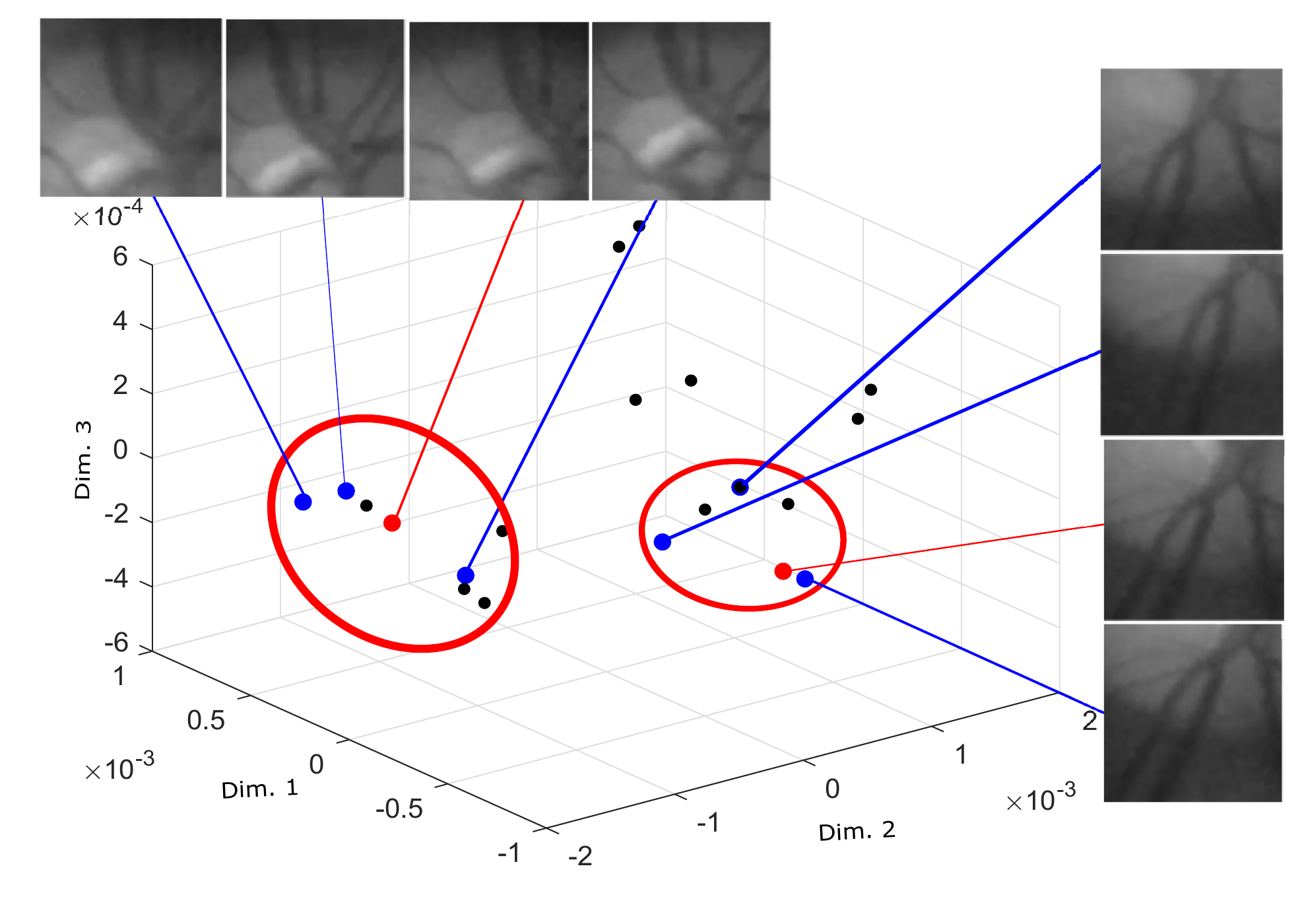}
	\caption{Features of images to be stitched in the top three dimensional space. Each small black dot indicate one mapped image. The colored dots in red circles show two selected samples (red) with their nearest three neighbors (blue). Please note that the distance is measured in the top 20 dimensional space.} 
	
	\label{fig:mosiacPCA}
\end{figure}

\begin{figure*}[!tp]
	\centering
	\includegraphics[width=1\textwidth]{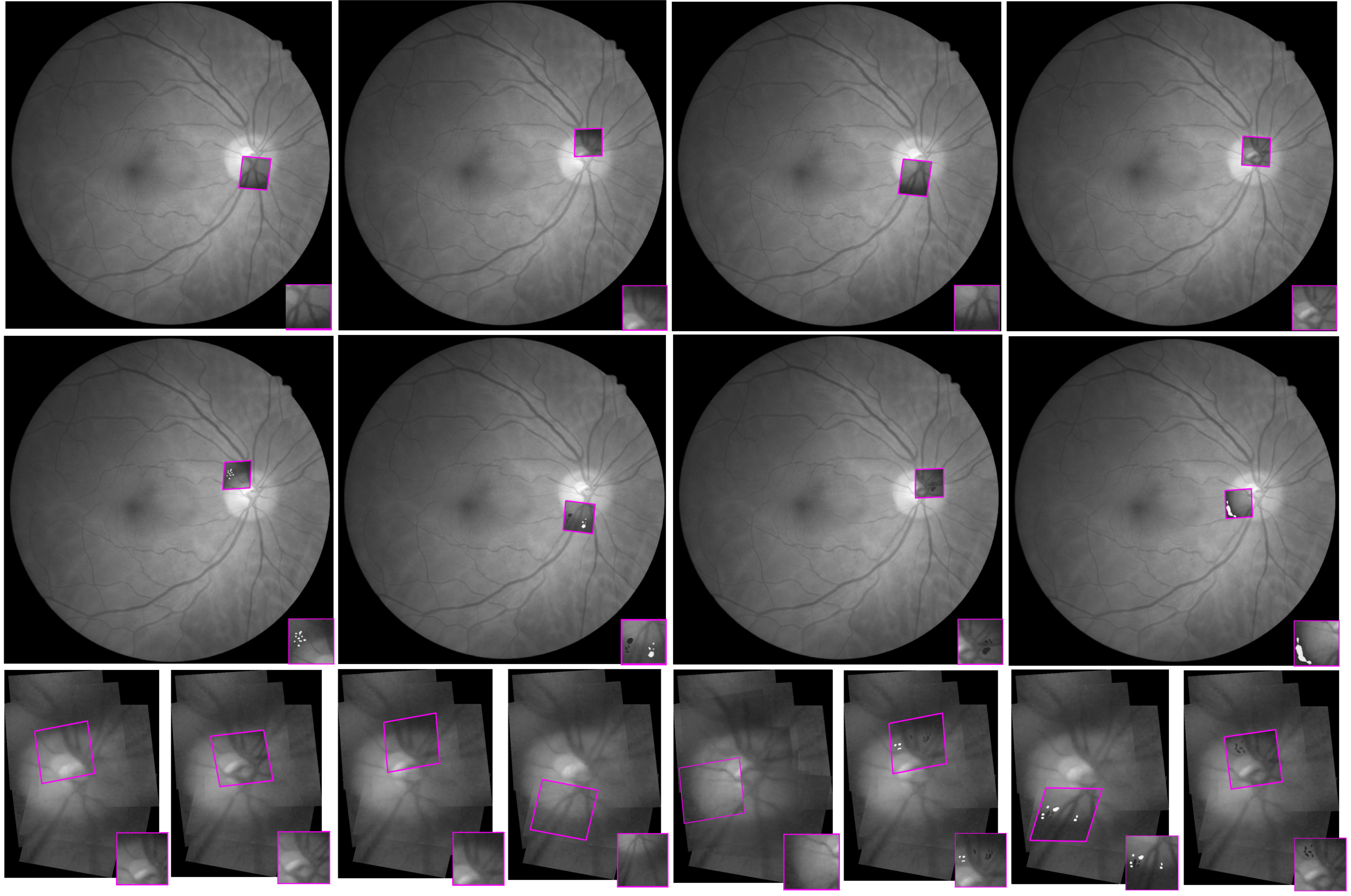}
	\caption{Examples of RetinaMatch results with and without artifacts. The first two rows are results of experiment 2 and the third row represents experiment 3. Each Template is shown at the right bottom corner. The mapped template on the full image covers the original area and is boxed with magenta lines.}
	\label{fig:results}
\end{figure*}

The full image in this experiment is mosaicked with 20 D-eye images using the proposed image stitching method. Based on no training for the D-eye user and other limitations of the procedure, we collected images covering the region around the optic disc. 
In the implementation, we used the first 20 dimensions of the features when computing the image distances in the low-dimensional space. Fig. \ref{fig:mosiacPCA} illustrates the distribution of the first three dimensions of the features. From the two examples in the figure, we can see the nearest three neighbors of the selected sample in the low dimensional space also have a large overlap in the image space. In the image patch registration, the MI-based registration method is applied. 
The last row of Fig. \ref{fig:results} shows the mosaicking result of the registration method with proposed method. The MI of the top three candidate neighbors are validated to be effective to choose the correct neighbor. The stitched full image has a 10$^\circ$ FOV with the same resolution as the D-eye templates. The image blending is not our focus and the mosaicked image is not perfect seamless.
  

\subsubsection{Template Matching}
Similar to experiment 2, we validate our method by matching 100 D-eye templates with and without pathological artificial features onto the mosaicked image. The images used for the mosaicking are not contained in the 100 template test set. The TRE results are shown in Table \ref{TRE_exp3}. The success rate is the percentage of matching pairs with TRE less than 8 pixels. 
RetinaMatch can match 96\% of image pairs without artifacts and 94\% of image pairs with artifacts. The TRE results were not much different from the observer variability. On the other hand, ASIFT cannot find the alignment position since the detected feature points are not sufficient for matching. The MI approach has a low rate of success as well, which has a high probability to cause mis-detection of emerging changes. 


\begin{table*}[tp]
	
	\centering
	\caption{Target registration error (TRE) of template matching methods in experiment 3.}
	\label{TRE_exp3}
	\setlength{\tabcolsep}{0.008\textwidth}{
	\begin{tabular}{|c|c|c|c|c|c|c|c|}
		\hline
	\multirow{2}{*}{}&
\multicolumn{2}{c|}{ASIFT}&\multicolumn{2}{c|}{ MI}&\multicolumn{2}{c|}{RetinaMatch}&{Observer}\cr\cline{2-7}
&Mean$\pm$SD&Success Rate&Mean$\pm$SD&Success Rate&Mean$\pm$SD&Success Rate&Variability\cr
		\hline
		\hline
		Without artificial features& NA&0 & 2.77$\pm$1.54& 15\%& 3.06$\pm$1.65& 96\%& 2.80$\pm$1.02\cr\hline
		With artificial features& NA& 0& 3.34$\pm$1.52&8\% & 3.24$\pm$1.75 & 94\%&2.80$\pm$1.02\cr\hline
		
		\hline
	\end{tabular}}
\end{table*}

\section{Conclusion and Discussion}
\label{sec:DISCUSSION}

We present a new template matching method, RetinaMatch, which can be used in remote retina health monitoring with affordable imaging devices. A PCA-based coarse localization method is proposed to provide a good initialization for the MI-based registration in the template matching. In this way, RetinaMatch can obtain an accurate affine transformation between the image pair with poor quality and a large FOV baseline.  
As demonstrated in the first simulation experiments, RetinaMatch does not handle templates with large affine deformations, with the success rate decreasing at level 4 and 5 as shown in Fig. \ref{fig:RM1}. 
The limitation to smaller rotation angle is physiologically based. The human eye has a limited range of torsional rotation with respect to the visual axis \cite{van1994torsional}.  
Importantly, the template image captured by adapter-based optics with general operation will not have a linear deformation exceeding the RetinaMatch limit, and therefore we can ignore the poor performance over the third-level affine degradation.
To our knowledge, this is the first report addressing template matching in retina images whose template contains unconstrained small retinal areas rather than a specific object. Further algorithm testing is needed on the smartphone or other low cost fundus imaging platforms as all current testing has been limited to a PC workstation. 

To validate RetinaMatch, experiments using both human datasets with simulation and \textit{in vivo} retina images from a case study were performed. Experiments with simulated datasets allowed evaluation of the accuracy and robustness of RetinaMatch to different levels and sequences of degradations. The \textit{in vivo} case study ensures that our method can be applied using a consumer product. It was observed that RetinaMatch provided superior performance under different image conditions over standard ASIFT and MI algorithms. The parameters, such as the offset $f=10, f'=5$ and the dimension $d=20$, are independent of various datasets in the implementation of RetinaMatch, which makes it easier to translate our method to other similar imaging device besides \textit{D-eye}. 

The evaluation of the RetinaMatch accuracy is difficult in experiment 2 and 3 without ground truth. Since the goal of template matching is providing accurate alignment for downstream analysis, we use TRE as the metric.  Compared with entropy based measures or the similarity measure itself, TRE measures the result intuitively in pixels and is independent of different regularization methods.

The remote monitoring of retina health with template matching is the first medical application to be proposed with RetinalMatch. Tele-ophthalmology is a promising application since many diseases are manifested at an early stage that are detectable with optical imaging of the retina. Because early stage retinal diseases do not present with symptoms, routine screening is important for early detection, which requires both high sensitivity and even higher specificity. 
The adapter-based optics and the digital cameras from smart phones provide an efficient and economic approach to capture retina images regularly at home.
The images of the current state can be mapped with RetinaMatch and then compared with the previous state. With regular screening, the process of lesion formation and therapeutic treatment can be monitored over time. In the experiment, \textit{D-eye} is chosen just as one low-cost example among many others with small FOV on undilated pupils \cite{panwar2016fundus}.
Similar fundus imaging techniques can also be implemented in emerging commercial VR, AR, and mixed reality headsets that will be widespread in the future.

There are different kinds of retina lesion that can be screened with portable fundus cameras. In the medical example of monitoring hypertension, the larger arteries constrict and the venous vessels enlarge in diameter \cite{Kawasaki2009}. For example, the larger blood vessel cross-sectional diameter is about 20 pixels in the case study, and a change with hypertension will be in the range of 10-60\%, so we are looking for over 2-pixel changes from baseline over time. The TRE is shown to be extremely low and most errors are below 2 pixels (excluding observer variability) in Tables \ref{TRE_exp2} and \ref{TRE_exp3}. With advanced trend analytics \cite{dai2017trend}, we can expect template matching errors to be well below a threshold of clinical significance. For more precise vessel width measurement, RetinaMatch can be combined with vessel segmentation, as described in our previous publication \cite{gong2018measurement}. The vessels of interest can be located on the current templates and the corresponding vessel width is then obtained by segmentation around the mapped location. Note the vessel segmentation here is applied on very small retina patches ($20\times20$ pixels), which is more robust and accurate than segmentation of wide FOV retina images. The segmentation error in \cite{gong2018measurement} is less than 1 pixel, which has been presented using \textit{D-eye} images. Xu et al. \cite{xu2016smartphone} proposed the vessel width segmentation and measurement on retina imaging acquired from the low quality fundus camera as well. They also report similar 1-pixel accuracy. However, the imaging device they used produced higher quality retinal images, having five times larger FOV than the \textit{D-eye}. The biomarkers of abusive head trauma (AHT) is another example. The most common retinal manifestation of AHT is multiple retinal hemorrhages in multiple layers of the retina \cite{yusuf2017non}. Matching the captured images onto the full retina image, The hemorrhagic spots can be easily segmented after the subtraction of the current retina regions and previous status. The AHT then can be recognized automatically when such spots are detected with portable fundus cameras.  

RetinaMatch may be used in other medical image applications for template matching. For example, in the case of endoscopic guidance of therapy by a surgical robot \cite{hu2018semi}, the current limited-sized FOV can be matched onto the panorama for endoscope localization. Thus, this image template matching technique can be used to create a more reliable closed-loop control for the robot arm and surgical tool guidance.

\bibliographystyle{IEEEtran}
\bibliography{IEEEref}

%

%
%
%




\end{document}